\documentclass[]{revtex4-2}
\usepackage[english]{babel}

\usepackage{hyperref}

\usepackage[letterpaper,top=2cm,bottom=2cm,left=3cm,right=3cm,marginparwidth=1.75cm]{geometry}

\usepackage{amsmath}
\usepackage{graphicx}

\begin{document}

\title{Friendship paradox disappears under degree biased network sampling}
\author{Wojciech Roga}
\email{wojciech.roga@keio.jp}
\affiliation{%
Department of Electronics and Electrical Engineering, Keio University, 3-14-1 Hiyoshi, Kohoku-ku, Yokohama 223-8522, Japan}

\begin{abstract}
We show that in an undirected graph under degree biased sampling the expected degree of vertices is equal to the expected degree of their neighbors. In consequence, under the biased sampling the social network result known as the friendship paradox disappears. The identity is equivalent to the existence of a stationary state of a random walk on the graph or to the conservation of the total flow defined by the difference of the degrees of the vertices. 
\end{abstract}

\maketitle

\section{Introduction}

The friendship paradox \cite{feld1991your} is a famous social network or graph theory statement. It says that the expectation value of a degree of a uniformly sampled vertex is not greater than the average degree of the network vertices' neighbors. Here, the last average is calculated by summing all degrees of the graph vertices' neighbors and dividing it by the number of the graph edges \cite{feld1991your}, or by taking the average locally weighted by the degree of an individual \cite{WOS:000750884500003}. However, the individuals, are again sampled uniformly from the graph. The phenomenon is also discussed with different statistics \cite{WOS:000333897200003}.

In the original work \cite{feld1991your} the interpretation was given in terms of the number of friends in a social network environment: "The mean number of friends of friends is always greater than the mean number of friends of individuals". 
The result was generalized in many ways leading to similar "paradoxical" conclusions regarding other characteristics of the individuals if they are positively correlated with the degree, such as the happiness, or the number of citations \cite{WOS:000402201200001, WOS:000333897200003}.

The friendship paradox reflexes a bias due to chosen statistics of samples. This bias may lead to majority illusions \cite{WOS:000371218400007,WOS:001564008400001} when individuals have the impression that their environment is representative of the total network, while it is not. Also, the sampling statistics chosen to investigate the network may lead to systematic errors in the context of medical surveys \cite{WOS:000333897200003}, distorted impressions and bad decisions in such contexts as finance \cite{WOS:000506809500002}, opinions of certain behaviors \cite{WOS:000463838300008}, or website page rank assignment \cite{WOS:000295341600005}.The different properties of the sampling and the graph that influence the bias stated in the paradox have been studied in many works \cite{lee2026friendship, WOS:001564008400001, WOS:000295341600005}. 
In addition, ways of eliminating the paradox, for example, by edge cutting \cite{WOS:001477539600031}, have been discussed.

In this paper, we look at the scenario with the average deviation of the degree of a vertex from the average of its neighbors' degree and observe that such average deviation disappears under sampling biased by the degree. This way of counting averages is linked to natural universal properties of finite graphs such as the existence of a steady state of the random walk, or the conservation of the total flow (specifically defined) in the graph. 

The degree sampling has been studied, for example, in \cite{lovasz1993random}. It was useful for tasks such as finding influential nodes \cite{WOS:000344157600002}. It is realized by statistics induced by the steady state of the random walk \cite{lovasz1993random, WOS:000295341600005}.  

Our central observation is that under the degree sampling the difference between the node degree and the average neighbor degree is zero and the friendship paradox vanishes. Noticing that the degree sampling and the considered difference lead to the universal conclusion implies that any other choice of sampling or definition of the local difference lead to systematic bias that may induce undesired overestimations and errors. 

Our main observation is simple and is based on standard techniques which appear in the context of the friendship paradox. However, we do not find it stated explicitly. The probable reason is that the degree sampling is not the preferable way to explore large graphs, it may imply multiple samples of the same nodes, so network researchers usually focus on other methods. Nevertheless, we think that it is important to state the observation explicitly as it clarifies under which definitions and sampling method the paradox does not appear, and by deviation from this, when the paradox is expected.  




\section{Degree biased sampling identity}

{\bf Observation 1.}
Consider an undirected graph. If you randomly choose vertices according to degree sampling,
the expected imbalance between the degree of that vertex and the average degree of the neighbors of that vertex is zero.\\

\noindent{\bf Proof:}
Denote $k_1, k_2, ..., k_N$ the degrees of vertices $1, 2, ..., N$.
The probability of picking vertex $i$ is
$$
p_i=\frac{k_i}{2|E|},
$$
where $|E|$ is the total number of edges and it holds that 
$$
2|E|=\sum_i k_i.
$$
Assume that the neighbors of $i$ belong to a set $S_i$. The average degree of neighbors of $i$ is
$$
\frac{\sum_{j \in S_i} k_j}{k_i}.
$$
The local imbalance is
$$
 \frac{\sum_{j \in S_i} k_j}{k_i} - k_i = \frac{1}{k_i} ( \sum_{j \in S_i} k_j - k_i^2).
$$
The degree biased expectation value of the imbalance is
$$
\sum_i \frac{k_i}{2|E|} \left[\frac{1}{k_i} \left( \sum_{j \in S_i} k_j - k_i^2\right)\right].
$$
But it is known that, 
$$
\sum_i \sum_{j \in S_i} k_j = \sum_i k_i^2.
$$
This is the core identity in the proof of the friendship paradox \cite{feld1991your}. It means that when you sum over all vertices' neighbors, the $i$-th vertex with degree $k_i$ appears as a neighbor of $k_i$ vertices. 

So, the expected value of the local imbalance is 0. This finishes the proof.

\section{Random walk}

Observation 1 has a clear interpretation in terms of a stationary state of the random walk on the graph \cite{lovasz1993random,levin2017markov,albert2002statistical}.
Let us define the walk as follows:
\begin{itemize}
\item start at a chosen vertex,
\item move to the next vertex chosen from the connected ones uniformly at random,
\item repeat the procedure forever.
\end{itemize}
After an appropriately long time, the probability of finding the walker in the vertex $i$ \cite{lovasz1993random} is 
$$
p_i=\frac{k_i}{2|E|}.
$$
The expected degree of the vertex in which the walker is currently is  
$$
d_{now}=\sum_i\frac{k_i^2}{2|E|}.
$$
The expected degree of the vertex in which the walker will be in the next step is
$$
d_{next}=\sum_i\frac{k_i}{2|E|}\sum_{j\in S_i}\frac{k_j}{k_i}=\frac{1}{2|E|}\sum_i\sum_{j\in S_i}k_j,
$$
and since the process is stationary 
$$
d_{now}=d_{next}.
$$
So, in terms of the random walk on the graph, Observation 1 is equivalent to the time invariance of the stationary state of the random walker on the undirected graph who for the next step chooses uniformly from all available options
$$
d_{next}-d_{now}=\sum_i p_i\left(\sum_{j\in S_i}\frac{k_j}{k_i}-k_i\right)=0.
$$

\section{Conservation of flow}

Yet another equivalent formulation of Observation 1 is the conservation of the total flow in the graph. 
Let us define for the edge $i\rightarrow j$ the flow given by the degree imbalance  \cite{farzam2020degree}
$$
\phi(i\rightarrow j)=k_i-k_j.
$$
For a vertex $i$, let us define net flow (divergence) as
$$
{\rm div}(i)=\sum_{j\in S_i}\phi(i\rightarrow j).
$$
We have
\begin{eqnarray*}
{\rm div}(i)=\sum_{k_j\in S_i}(k_i-k_j)
=k_i^2-\sum_{k_j\in S_i}k_j.
\end{eqnarray*}
Hence, Observation 1 is equivalent to 
$$
\sum_i{\rm div}(i)=0.
$$
Which is the law of conservation of the total flow in the network. 

\section{Simulations}

We run simulations of the random walk on three graphs of different size and structure: random Erdős–Rényi graph, Zachary'd Karate Club graph \cite{zachary1977information}, and SNAP Facebook graph \cite{McAuley2012LearningTD}. In each step we checked the local imbalance and take the uniform average of the local imbalances. We plot the running averages together with the standard error of the mean (SEM). In all cases the average converges to zero, while the SEM converges to a constant value depending on the network.  

\begin{figure}
\centering
\includegraphics[width=1\linewidth]{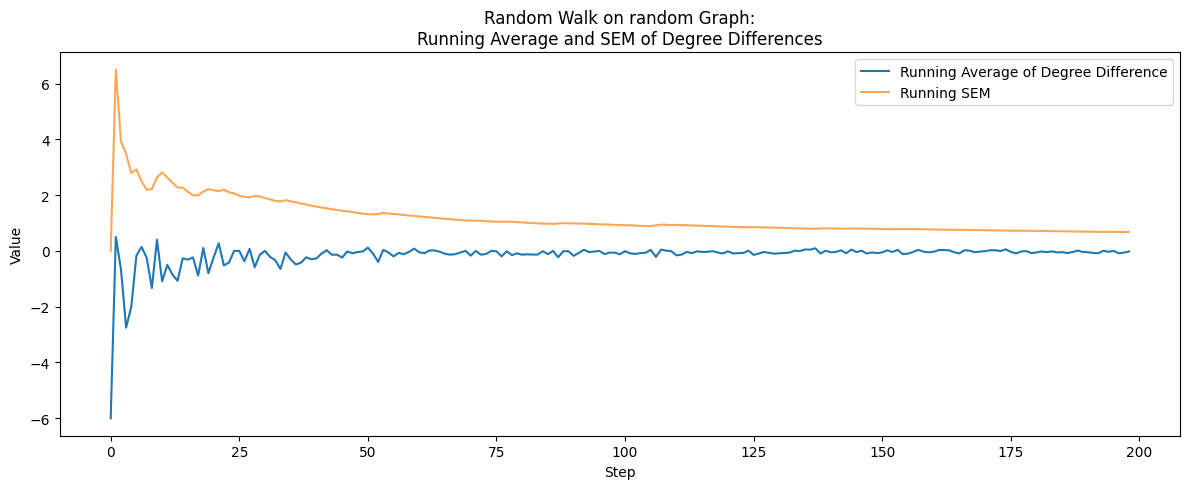}
\caption{\label{fig:erdos} Random walk on random graph (Erdős–Rényi) with $n=1000$ nodes and probability of each pair of nodes being connected $p=0.05$. }
\end{figure}

\begin{figure}
\centering
\includegraphics[width=1\linewidth]{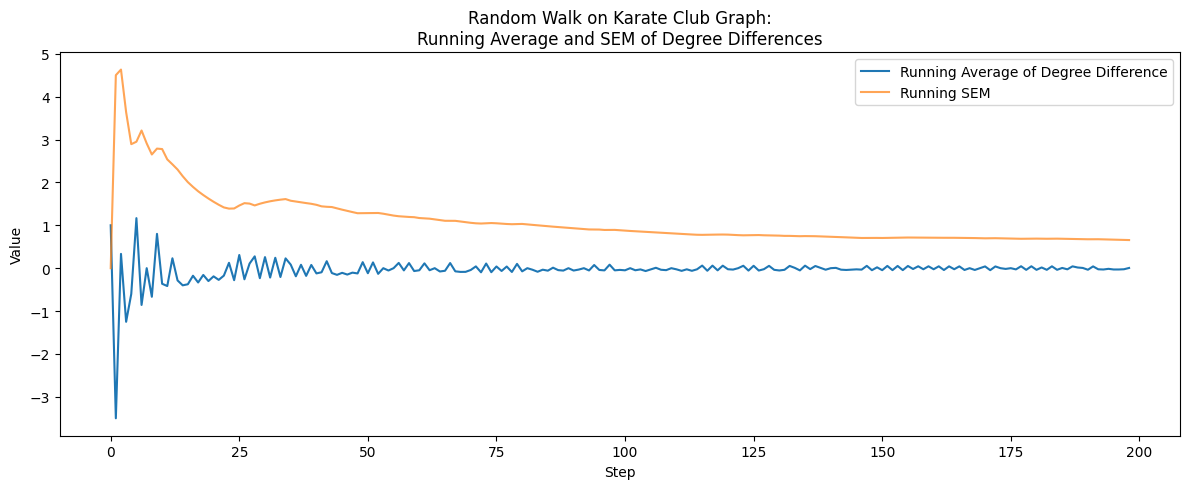}
\caption{\label{fig:karate}Random walk on Zachary's Karate Club graph with 34 nodes and 78 edges.}
\end{figure}

\begin{figure}
\centering
\includegraphics[width=1\linewidth]{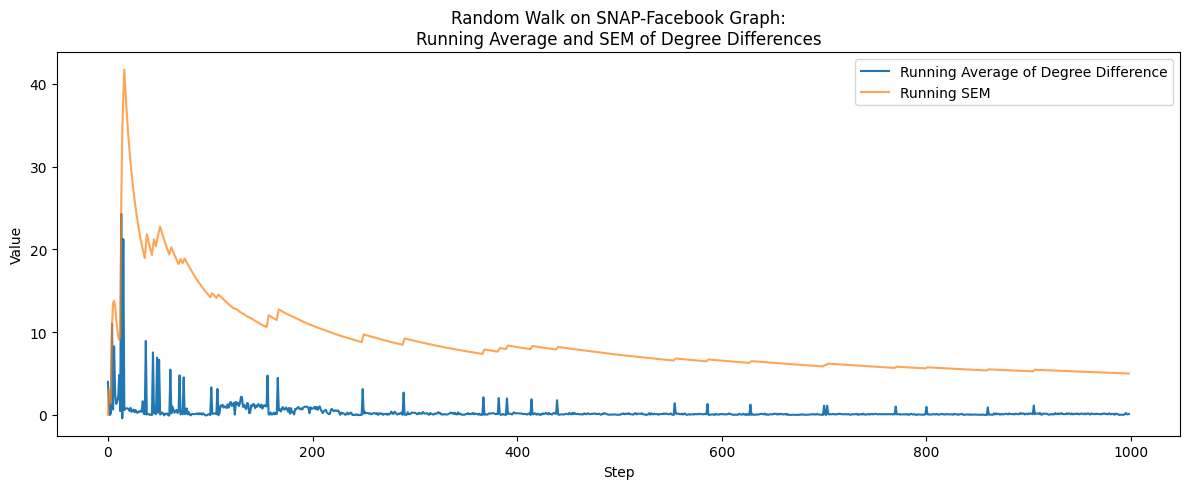}
\caption{\label{fig:facebook}Random walk on SNAP Facebook graph (${\rm facebook\_combined}$) with 4039 nodes and 88234 connections.}
\end{figure}

The first example is the Erdős–Rényi graph $G(n,p)$ with $n=1000$ nodes and the probability that any two nodes are connected $p=0.05$. The results are shown in Fig. \ref{fig:erdos}.
The second example refers to the Zachary's Karate Club graph \cite{zachary1977information} with 34 nodes and 78 edges. The results are shown in Fig. \ref{fig:karate}.
As the third example, we consider the Stanford SNAP Facebook dataset \cite{McAuley2012LearningTD} with 4039 nodes and 88234 edges. The results are shown in Fig. \ref{fig:facebook}.

In all cases, the expectation value converges to zero in agreement with Observation 1. We observe that for large graphs this happens in a relatively small number of steps, before the walker explores the total graph.

\section{Discussion}

The key identity used in the friendship paradox
$$
\sum_i\sum_{j\in S_i}k_j=\sum_ik_i^2
$$
is the key identity also in our case. In the friendship paradox, the average number of friends of friends is this quantity divided by $2|E|=\sum_i k_i$, which is shown to be larger than the uniformly averaged degree of vertices \cite{feld1991your}
$$
\frac{\sum_i\sum_{j\in S_i}k_j}{\sum_i k_i}\geq \frac{\sum_i k_i}{N}.
$$ 
In our case, on the right hand side, we have the average biased by the degree that leads to $\sum_ik_i^2/\sum_i k_i$. In our case, the left hand side contains the local average $\sum_{j\in S_i}k_j/k_i$ which then is averaged over the degree biased distribution $k_i/\sum_i k_i$. So instead of a global average of the number of friends' friends, we use the local average. Instead of taking a uniform average over the friends of individuals, we use the connection bias. Then the inequality and the friendship paradox disappear.

One can look at this from the perspective of the internet crawling robot (network random surfer \cite{grady2010discrete}). While randomly walking through the network, instead of randomly picking the vertices, the robot sees the imbalance between the number of friends and the average number of friends' friends only locally. Such robot does not observe the friendship paradox. The expected number of friends is the same as the expected number of friends of friends.

The random walk type of sampling is not preferable by network researches. For large network it converges slowly to the steady state and it visited the same nodes many times. Nevertheless, our observation is crucial to realize that deviation from the random walking type of averaging of the local degree imbalance defined in this paper naturally leads to biases such as the friendship paradox.

\section*{Acknowledgment}
This work was supported by JST Moonshot R\&D Grant No. JPMJMS2061, JST ASPIRE, Grant No. JPMJAP2427, JST COI-NEXT Grant No. JPMJPF2221.

\bibliographystyle{IEEEtran}
\bibliography{sample3}

\end{document}